\documentclass[aps,pra,twocolumn,amsmath,nofootinbib]{revtex4-1}
\usepackage{epsfig}
\usepackage[colorlinks,linkcolor=blue,anchorcolor=blue,citecolor=blue,urlcolor=blue,breaklinks=true]{hyperref}
\usepackage{graphics}
\usepackage{color}
\usepackage{epstopdf}
\usepackage{bm}
\usepackage{graphicx}
\usepackage{dcolumn}
\usepackage{cases}
\usepackage{appendix}
\usepackage{ulem}
\usepackage{subfigure}

\begin{document}

\author{Meng-Yun Lai$^{1,2}$}~\email[]{mengyunlai@gmail.com}
\author{Yong-Long Wang$^{1,2}$}~\email[]{wangyonglong@lyu.edu.cn}
\author{Guo-Hua Liang$^{3}$}~\email[]{guohua@nju.edu.cn}
\author{Hong-Shi Zong$^{1,4,5}$}~\email[]{zonghs@nju.edu.cn}
\address{$^{1}$Department of Physics, Nanjing University, Nanjing 210093, China}
\address{$^{2}$School of Physics and Electronic Engineering, Linyi University, Linyi, 276005, China}
\address{$^{3}$College of engineering and applied sciences, Nanjing University, Nanjing 210093, China}
\address{$^{4}$Joint Center for Particle, Nuclear Physics and Cosmology, Nanjing 210093, China}
\address{$^{5}$State Key Laboratory of Theoretical Physics, Institute of Theoretical Physics, CAS, Beijing, 100190, China}


\title{Geometrical phase and Hall effect associated with the transverse spin of light}
\begin{abstract}
By analyzing the vectorial Helmholtz equation within the thin-layer approach, we find that light acquires a novel geometrical phase, in addition to the usual one (the optical Berry phase), during the propagation along a curved path. Unlike the optical Berry phase, the novel geometrical phase is induced by the transverse spin along the binormal direction and associated with the curvature of the curve. Furthermore, we show a novel Hall effect of light induced by the torsion of the curve and associated with the transverse spin along the binormal direction, which is different from the usual spin Hall effect of light. Finally, we demonstrate that the usual and novel geometrical phase phenomena are described by different geometry-induced U(1) gauge fields in different adiabatic approximations. In the nonadiabatic case, these gauge fields are united in one effective equation by SO(3) group.
\end{abstract}

\maketitle

\section{Introduction}
It has long been known that light carries the spin angular momentum (AM) along the direction of propagation. The longitudinal spin is generated by rotating electric and magnetic fields in the transverse plane. In the past decade, it was shown that light exhibits an unusual transverse spin in various structured optical fields such as evanescent waves, interference fields and focused beams \cite{Bliokh2014NC,Aiello2015,BLIOKH2015physrep}. The transverse spin of light is associated with the presence of a nonvanishing longitudinal component of the electromagnetic field. As early as 1959, Richards and Wolf showed that in an optical focusing system the electromagnetic field oscillates along both transverse and longitudinal directions with $\pi/2$ phase difference \cite{Richards1959}. However, they did not identify this elliptical polarization in the propagation plane as the manifestation of nonzero transverse spin of light. Only in 2009 the transverse spin of light and its extraordinary properties, which are very different from the longitudinal one, started to attract researchers' interests \cite{Aiello2009,Aiello2010}. Most importantly for applications, the transverse spin is locked to the direction of propagation: the sign of the transverse spin AM density flips when the propagation direction reverses \cite{Marrucci2014}. Interestingly, the spin-direction locking in evanescent waves can be recognized as the optical counterpart of the quantum spin Hall effect \cite{Bliokh1448}. Owing to this robust spin-direction locking, the transverse spin AM has found important applications in highly efficient unidirectional optical transport \cite{Rodriguez-Fortuno2013,Miroshnichenko2013,Junge2013,Petersen67,Mitsch2014,O’Connor2014,Marrucci2014,Pichler2015}, resulting in a young yet advanced research field: the chiral quantum optics \cite{Lodahl2017}.

On the other hand, light acquires an optical Berry phase when propagates along a curved path \cite{Ross1984,Chiao1986,Tomita1986,Haldane:86,berry1987interpreting}. The optical Berry phase underpins the spin-orbit interactions (SOI) of light which play a fundamental role in modern optics \cite{Bliokh2015Natpho}. The optical Berry phase can be easily understood through the following argument. Consider the three-dimensional vector space attached to each point of a space curve. Due to the vectorial nature of the light field, the three-dimensional vector space relates to the spin space of light propagating along the curve. For transverse electric field, the three-dimensional vector space reduces to a two-dimensional spin space of light. Consequently, the torsion of the curve, which describes the transport of the spin space along the curve and relates to the connection of the spin space, is coupled with the longitudinal spin and acts as a gauge field within the dynamics along the curve \cite{Lai2018PRA}. Then akin to the Aharonov-Bohm phase, this effective gauge field results in the optical Berry phase.

Along these lines it is naturally to expect that there are geometrical phase phenomena associated with the transverse spin of light \cite{Shao2018}, just like the longitudinal one. In fact, the geometrical quantity of the curve (i.e., the curvature) may be coupled with the spin AM of light. As the consequence, a novel class of geometrical phase phenomena, which is associated with the transverse spin of light, would take place. To study the transverse-spin-dependent geometrical phase phenomena, in Sec.~\ref{effectiveEq} of this paper we use the thin-layer approach to analyze the vectorial Helmholtz equation and take into consideration the longitudinal component of the electric field at the same time. The thin-layer approach is a convenient framework to study the evolution of various types of waves, including the electromagnetic wave, along a curve or on a curved surface \cite{Costa1981,burgess1993,Ouyang1999,batz2008,Schultheiss2010,TAIRA2011,yong-longwang2014,yong-longWang2017,liang2018pra}.
In Sec.~\ref{geometricalEffects}, we discuss the transverse-spin-dependent geometrical phase phenomena. In Sec.~\ref{GaugeStructure}, we analyze the gauge structure of the present theory. Section \ref{conclusions} provides conclusions.

\section{Effective equation}\label{effectiveEq}
The vectorial Helmholtz equation describing the propagation of the electromagnetic wave is obtained from the combination of Maxwell's equations \cite{batz2008,Lai2018PRA},
\begin{eqnarray}\label{vectorialHelmholtzEq}
  \nabla_j\nabla^j E^i -\frac{n^2}{c^2}\partial_t^2{E^i} &=& 0.
\end{eqnarray}
According to the well-known result in differential geometry \cite{nakahara2003geometry}, the covariant derivative $\nabla_i$ is different from the ordinary derivative $\partial_i=\partial/\partial q^i$ when acts on a tensor (including vector). For example, when $\nabla_i$  acts on a second order tensor $T^{jk}$, we have
\begin{eqnarray}
  \nabla_i T^{jk} &=& \partial_i T^{jk} +\Gamma^j_{il}T^{lk}+\Gamma^k_{il}T^{jl},
\end{eqnarray}
where $\Gamma^i_{jk}$ is the Christoffel symbol which relates to the parallel transport of tensors. Apparently, the Christoffel symbols arise in the Laplacian term, $\nabla_j\nabla^j E^i$, in the vectorial Helmholtz equation (\ref{vectorialHelmholtzEq}):
\begin{eqnarray}\label{LaplacianTerm}
  \nabla_j\nabla^j E^i &=& \frac{1}{\sqrt{g}}\partial_j(\sqrt{g}g^{jk}\partial_kE^i)+2g^{jk}\Gamma^i_{kl}\partial_jE^l
  \nonumber\\
  &&+\frac{1}{\sqrt{g}}\partial_j(\sqrt{g}g^{jk}\Gamma^i_{kl})E^l+g^{jl}\Gamma^i_{jk}
  \Gamma^k_{lm}E^m,
\end{eqnarray}
where $g_{ij}$ denotes the metric tensor and $g=det(g_{ij})$. For comparison, the usual Helmholtz equation is
\begin{eqnarray}\label{HelmholtzEq}
\nabla_j\nabla^j E -\frac{n^2}{c^2}\partial_t^2{E} &=& 0,
\end{eqnarray}
where $\nabla_i\nabla^i E=1/{\sqrt{g}}\partial_j(\sqrt{g}g^{jk}\partial_kE)$ and $E$ denotes the scalar approximation for the electric field. Note that the first term on the right hand side of Eq.~(\ref{LaplacianTerm}) is exactly the Laplacian term in Eq.~(\ref{HelmholtzEq}). This term can be expanded as
\begin{eqnarray}
  \frac{1}{\sqrt{g}}\partial_j(\sqrt{g}g^{jk}\partial_kE^i) &=& (\partial_j\partial^j +\Gamma^l_{lj}\partial^j)E^i
\end{eqnarray}
where the last term leads to the effective gauge field associated with the intrinsic orbital AM \cite{SCHUSTER2003132,TAIRA2011,Lai2018PRA,wang2018PRA}. Unlike this pure orbital term, the last three terms in Eq.~(\ref{LaplacianTerm}) only appear in the case of vector waves and, thus, exhibit the vectorial properties of the electromagnetic waves.  The second and third terms on the right hand side of Eq.~(\ref{LaplacianTerm}) represents the geometry-induced SOI of light, since they mix the spin and orbital degrees of freedom by index contraction. The last term in Eq.~(\ref{LaplacianTerm}) is a pure spin term, only relates to the spin degree of freedom. Physically, these geometry-induced terms are associated with the change of the direction of the wavevector of light. In practice, the variation of the wavevector direction can be achieved by constructing various optical systems, such as optical fibers \cite{Chiao1986}, two dimensional curved waveguides \cite{batz2008}, gradient-index media \cite{Bliokh2009JOA,Bliokh2009PRA} and plasmonic nanostructures \cite{haefner2009PRL,pan2016PRA}.

\begin{figure}
  \centering
  \includegraphics[width=0.25\textwidth]{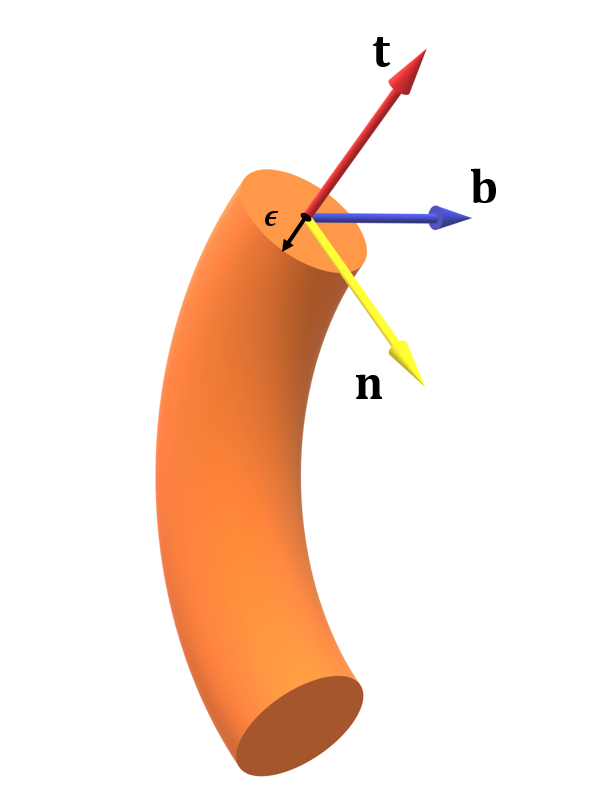}
  \caption{Schematic of an imaginary curved thin tube, $\epsilon$ is the radius of the normal cross section.}\label{001}
\end{figure}

\begin{figure}
\centering
\subfigure[]{ \label{002a} \includegraphics[width=1.55in]{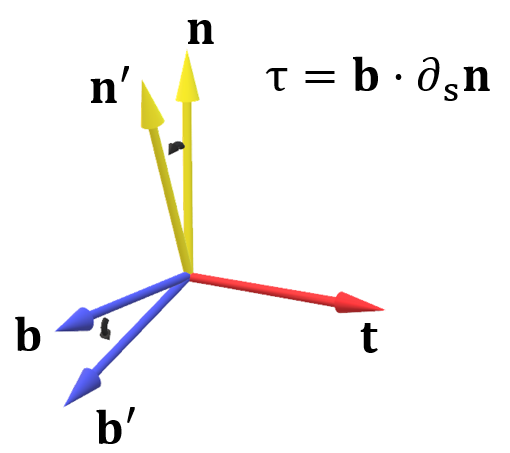}} \hspace{0.1in} \subfigure[]{ \label{002b}
\includegraphics[width=1.55in]{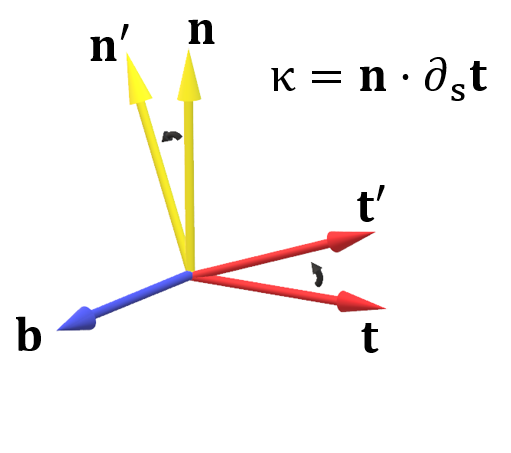}} \caption{Curvature and torsion as the components of an angular velocity vector.} \label{002}
\end{figure}

Here, we consider a heuristic model to show how the geometrical quantities influence the propagation of light carrying transverse spin. We focus on the curved propagation of an electromagnetic wave and take into consideration the longitudinal component of the electric field. First of all, let us introduce a coordinate system (we prefer to call it ``adapted coordinate system'') \cite{SCHUSTER2003132}, in which case the present problem can be simply described. Let ${\bf r}_0(s)$ be the parametrization of the curve along which the electromagnetic wave propagates, with $s$ being the arc-length. In the vicinity of ${\bf r}_0(s)$, the points are described by the following position vector (see Fig.~\ref{001}):
\begin{eqnarray}\label{PositionVector}
  {\bf R}(s,q^2,q^3) &=& {\bf r}_0(s)+q^2{\bf n}(s)+q^3{\bf b}(s),
\end{eqnarray}
where ${\bf n}$ and ${\bf b}$ are the unit normal and binormal vectors of ${\bf r}_0(s)$, and $q^2, q^3$ are the corresponding coordinate variables. By introducing the position vector, we define the adapted coordinates system in the vicinity of ${\bf r}_0(s)$. The metric tensor in this coordinates system is given by
\begin{eqnarray}\label{Metric0}
  g_{ij} &=& \partial_i{\bf R}\cdot\partial_j{\bf R}.
\end{eqnarray}
Note that the unit tangent, normal and binormal vectors $\{{\bf t}=\partial_s{\bf r}_0(s), {\bf n}, {\bf b}\}$ construct the Frenet frame of ${\bf r}_0(s)$. According to the Frenet-Serret formulas, these three vectors and their derivatives with respect to the arc-length obey the following equation:
\begin{eqnarray}\label{F-Sformulas}
  \left(
     \begin{array}{c}
       \partial_s {{\bf t}}\\
       \partial_s {{\bf n}} \\
       \partial_s {{\bf b}} \\
     \end{array}
   \right)
   &=& \left(
          \begin{array}{ccc}
            0 & \kappa(s) & 0 \\
            -\kappa(s) & 0 & \tau(s) \\
            0 & -\tau(s) & 0 \\
          \end{array}
        \right)
        \left(
                 \begin{array}{c}
                   {{\bf t}} \\
                   {{\bf n}} \\
                   {{\bf b}} \\
                 \end{array}
               \right),
\end{eqnarray}
where $\kappa(s)$ and $\tau(s)$ are the curvature and torsion of ${\bf r}_0(s)$ which describe the rotations of ${\bf t}-{\bf n}$ and ${\bf n}-{\bf b}$ planes along the curve, respectively (see Fig. \ref{002}).  Substituting the position vector ($\ref{PositionVector}$) and Frenet-Serret formulas (\ref{F-Sformulas}) into the metric (\ref{Metric0}), we calculate the metric as:
\begin{eqnarray}\label{metric}
  g_{ij} &=& \frac{\partial {{\bf R}}}{\partial q^i}\cdot\frac{\partial {{\bf R}}}{\partial q^j} \nonumber\\
  &=& \left(
  \begin{array}{ccc}
   g_{11} & -\tau(s)q^3 & \tau(s)q^2 \\
  -\tau(s)q^3 & 1 & 0 \\
  \tau(s)q^2 & 0 & 1 \\
  \end{array}
  \right),
\end{eqnarray}
where $g_{11}=[1-\kappa(s)q^2]^2+\tau(s)^2[(q^2)^2+(q^3)^2]$.

Naively speaking, by using the adapted coordinate system we assume that the electromagnetic wave propagates in an imaginary curved thin tube in which its axis is ${\bf r}_0(s)$ (see Fig.~\ref{001}). The shape of the cross section of the curved thin tube represents the normal symmetry of the system which we study in specific problem, and can be completely different in different problems. In the spirit of the thin-layer approach, the shape of the cross section determines the symmetry of the normal part of the wavefunction and thus some properties of the tangent dynamics \cite{wang2018PRA}. In the present paper, however, we are mostly concerned about the geometrical phase phenomena associated with the spin AM of the electromagnetic wave. The discussion of the shapes of the cross section is left out in the present paper, and we simply assume that the shape of the thin tube is circular (see Fig.~\ref{001}).

Following the thin-layer procedure discussed in Refs.~\cite{SCHUSTER2003132,Lai2018PRA,wang2018PRA}, we introduce a new wave function $\bar{E}^i=g^{1/4}E^i$, where $g^{1/4}$ is a rescale factor determined by the correct volume measure on ${\bf r}_0(s)$. Then substitute this rescaled wave function and the metric (\ref{metric}) into the vectorial Helmholtz equation (\ref{HelmholtzEq}). At last, integrate over the normal cross section and impose the thin-layer limiting $\epsilon\rightarrow 0$ to Eq.~(\ref{HelmholtzEq}). By doing so, we obtain the effective equation describing the electric field propagating along the curve ${\bf r}_0(s)$, that is:
\begin{eqnarray}\label{tangentEq}
  \big[(\partial_s-i\tau\hat{S}_1-i\kappa\hat{S}_3)^2+\frac{\kappa^2}{4}+k^2_s\big]
\left(
  \begin{array}{c}
    \bar{E}^1_s(s) \\
    \bar{E}^2_s(s) \\
    \bar{E}^3_s(s) \\
  \end{array}
\right)
   &=& 0,
\end{eqnarray}
where $\hat{S}_1$ and $\hat{S}_3$ are two of the spin-1 matrices,
\begin{eqnarray}
  \hat{S}_1 = \left(
                  \begin{array}{ccc}
                    0 & 0 & 0 \\
                    0 & 0 & -i \\
                    0 & i & 0 \\
                  \end{array}
                \right),
  ~~~~~~~~~
  \hat{S}_3 =
   \left(
  \begin{array}{ccc}
  0 & -i & 0 \\
   i & 0 & 0 \\
   0 & 0 & 0 \\
  \end{array}
  \right),
\end{eqnarray}
$k_s$ is the wave number. Here, we have assumed the separation $\bar{E}^i(s,q^2,q^3)=\bar{E}^i_s(s)\bar{E}^i_{\bot}(q^2,q^3)$. $\bar{E}^i_s$ $(i=1,2,3)$ represent the tangent parts of the rescaled electric fields that along ${\bf t}$, ${\bf n}$ and ${\bf b}$ respectively. The effective equation (\ref{tangentEq}) is the key result in the present paper. It describes the curved propagation of an electromagnetic wave with nonvanishing longitudinal component. Comparing with the trivial one-dimensional wave equation, $(\partial_s^2+k_s^2){\bf E}=0$, there are three additional terms appear in Eq.~(\ref{tangentEq}). Unlike the scalar geometrical field $\kappa^2/4$, which results from the action of normal derivatives on the rescale factor \cite{yong-longWang2017}, the other two additional terms $-i\tau\hat{S}_1$ and $-i\kappa\hat{S}_3$ relate to the parallel transport of the electric field vector along a curve and act as effective gauge terms within the effective dynamics. These two effective gauge terms result in a number of geometry-induced SOI of light. The torsion-induced one represents the SOI related to the longitudinal spin of light, and thus is responsible for the optical Berry phase and the spin Hall effect associated with the longitudinal spin of light \cite{Chiao1986,Bliokh2008}. The other term is induced by the curvature, and associated with the transverse spin in the binormal direction. Therefore, we naturally expect that this curvature-induced effective gauge term would lead to transverse-spin-dependent geometrical phase phenomena. In addition,
note that expanding $(\partial_s-i\tau\hat{S}_1-i\kappa\hat{S}_3)^2$ in Eq.~(\ref{tangentEq}) would lead to the following term
\begin{eqnarray}
 -\kappa^2\left(
            \begin{array}{ccc}
              1 & 0 & 0 \\
              0 & 1 & 0 \\
              0 & 0 & 0 \\
            \end{array}
          \right).
\end{eqnarray}
This term is responsible for the nonadiabatic polarization changes \cite{berry1987interpreting,Lai2018PRA}. As we discuss in Sec.~\ref{GaugeStructure}, according to the spin gauge field theory \cite{BLIOKH200513}, this term is recognized as the off-diagonal term which is responsible for the transitions between different spin levels of light and can be neglected in the adiabatic approximation. Moreover, the main equations in Refs.~\cite{berry1987interpreting,Lai2018PRA} can be derived from Eq. (\ref{tangentEq}).

\section{Geometrical effects}\label{geometricalEffects}
The optical Berry phase and the corresponding spin Hall effect have been extensively studied for decades and are thus mostly left out in this paper. In the rest of the present paper, we focus on the curvature-induced effective gauge term and the corresponding geometrical phase phenomena.
\subsection{Transverse-spin-dependent geometrical phase}

\begin{figure}
  \centering
  \includegraphics[width=0.35\textwidth]{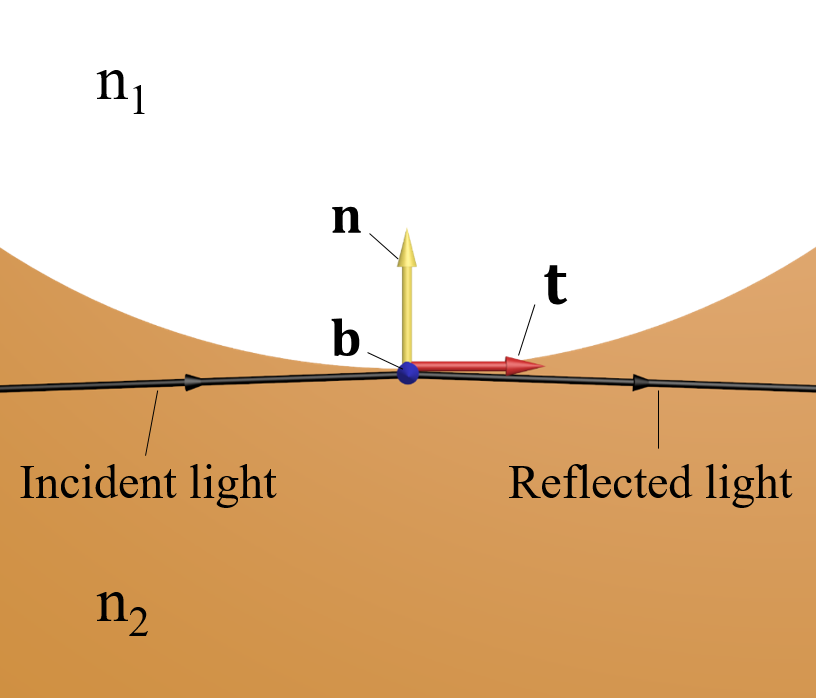}
  \caption{Total internal reflection at a curved interface.}\label{003}
\end{figure}


Without loss of generality, we consider a planar curve here, which means vanishing torsion. The effective equation (\ref{tangentEq}) becomes
\begin{eqnarray}\label{tangentEqT=0}
    \big[(\partial_s-i\kappa\hat{S}_3)^2+\frac{\kappa^2}{4}+k^2_s\big]
\left(
  \begin{array}{c}
    \bar{E}^1_s(s) \\
    \bar{E}^2_s(s) \\
    \bar{E}^3_s(s) \\
  \end{array}
\right)
   &=& 0.
\end{eqnarray}
Let us diagonalize the above equation by introducing the transverse circular polarizations \cite{Junge2013}
\begin{eqnarray}\label{Diagonalization}
   \left(
    \begin{array}{c}
      \bar{E}^{+} \\
      \bar{E}^{-} \\
      \bar{E}^3 \\
    \end{array}
  \right)
   &=&\frac{1}{\sqrt{2}}\left(
        \begin{array}{ccc}
          1 & -i &0\\
          1 & i &0\\
          0 &0 & \sqrt{2} \\
        \end{array}
      \right)
    \left(
         \begin{array}{c}
     \bar{E}^1 \\
      \bar{E}^2 \\
      \bar{E}^3 \\
         \end{array}
       \right).
\end{eqnarray}
As the result, Eq. (\ref{tangentEqT=0}) becomes
\begin{eqnarray}\label{tangentEqT=0,1}
    &&\big[(\partial_s-i\kappa\hat{\sigma})^2+\frac{\kappa^2}{4}+k^2_s\big]
\left(
  \begin{array}{c}
    \bar{E}^{+}_s \\
      \bar{E}^{-}_s \\
  \end{array}
\right)
   = 0, \\
&&(\partial_s^2+\frac{\kappa^2}{4}+k^2_s) \bar{E}^3_s = 0,
\end{eqnarray}
where $\hat{\sigma}=\left(
                      \begin{array}{cc}
                        1 & 0 \\
                        0 & -1 \\
                      \end{array}
                    \right)$.
One can readily find out that Eq. (\ref{tangentEqT=0,1}) describes a curvature-induced geometrical phase (transverse-spin-dependent geometrical phase)
\begin{eqnarray}\label{transversePhase}
  \Phi_{\perp} &=& {\sigma_{\perp}\int\kappa ds},
\end{eqnarray}
where $\sigma_{\perp}=\pm 1$ for $\bar{E}^{+}_s$ and $\bar{E}^{-}_s$ respectively, the subscript ``$\perp$'' indicates that the quantity is associated with the transverse spin of light. As the torsion-induced geometrical phase $\Phi_{\parallel}={\sigma_{\parallel}\int\tau ds}$ induces optical activity in the transverse plane \cite{Chiao1986}, the curvature-induced geometrical phase Eq. (\ref{transversePhase}) leads to an optical rotation in ${\bf t}-{\bf n}$ plane. Here, $\sigma_{\parallel}=\pm 1$ for left- and right-handed circularly polarizations, respectively \cite{Bliokh2009JOA,Bliokh2015Natpho}, and the subscript ``$\parallel$'' indicates that the quantity is associated with the longitudinal spin of light. In fact, these two geometrical phases together complete the parallel transport of the electric field with longitudinal component. Most importantly, the transverse-spin-dependent geometrical phase underpins a novel class of SOI effects of light. Actually, it has already been demonstrated experimentally that the transverse-spin-dependent geometrical phase results in the interaction between the intrinsic orbital AM and the transverse spin of light \cite{Shao2018}.

So far, we have discussed the transverse-spin-dependent geometrical phase of light from a rather theoretical viewpoint. Next we provide a heuristic but practical example to illustrate the influence of the geometrical phase on the curved propagation of an electromagnetic wave carrying transverse spin. The transverse spin AM of light arises in various structured optical fields such as evanescent waves, interference fields and focused beams \cite{Bliokh2014NC,Aiello2015,BLIOKH2015physrep}, due to the strong light confinement \cite{Lodahl2017}.
Here, we consider the evanescent wave generated by the total internal reflection at a curved interface between two different media (as a schematic, see Fig.~\ref{003}). We assume that the evanescent wave propagates along the curved interface and the total internal reflection is located at $s=0$. Experimentally speaking, the curved propagation of the evanescent wave may be realized by a whispering-gallery-mode (WGM) microresonator \cite{Junge2013,Shao2018} or a curved nanophotonic waveguide \cite{Petersen67,Lodahl2017}. For a planar interface located at $x=0$, the evanescent wave propagating along the $z$-direction can be written as \cite{Bliokh2014NC}
\begin{eqnarray}\label{evanescent0}
  {\bf E}_{p} &\propto& ({\bf e}_x-i\frac{k_x}{k_z}{\bf e}_z)e^{ik_zz-k_x x},
\end{eqnarray}
where ${\bf e}_x$ and ${\bf e}_z$ denote the basis vectors of Cartesian coordinate system, $k_z$ is the wave number and $k_x$ is the decrement.
Then, it would appear reasonable to generalize the evanescent wave (\ref{evanescent0}) to the curved case by demanding the propagation along the curved interface and decay along the perpendicular direction. Thus, the propagation of an evanescent wave along a curve is given by
\begin{eqnarray}\label{evanescent1}
  {\bf E}_c &\propto& [(A_+e^{i\int\kappa ds}+A_-e^{-i\int\kappa ds}){\bf t} \nonumber\\
  &&+i(A_+e^{i\int\kappa ds}-A_-e^{-i\int\kappa ds}){\bf n}]e^{ik_ss-k_n q^2},
\end{eqnarray}
where the transverse-spin-dependent geometrical phase is included, $A_+$ and $A_-$ are the initial amplitudes of the two opposite transverse circular polarizations. The ratio between $A_+$ and $A_-$ is determined by the initial condition ${\bf E}_c(s=0)\propto ({\bf n}-ik_n/k_s{\bf t})$. Substituting Eq.~(\ref{evanescent1}) into the spin AM density \cite{BLIOKH2015physrep}
\begin{eqnarray}
  {\bf S} &\propto& {\rm Im}({\bf E}^{*}\times{\bf E}),
\end{eqnarray}
one can readily find that the result is exactly the same as the case of planar interface. In other words, the geometrical phase do not change the transverse spin AM density. This suggests the robustness of the spin-direction locking in the case of curved interface. This trivial result is what we expect, because the transverse-spin-dependent geometrical phase rotates the field vector in the ${\bf t}-{\bf n}$ plane without changing its magnitude. Moreover, as a nontrivial result, one can directly verify that the ratio between the longitudinal and transverse components of the electric field vary periodically along the curved path, rather than being constant \cite{Junge2013}.

\begin{figure}
\centering
\subfigure[]{ \label{004a} \includegraphics[width=1.6in]{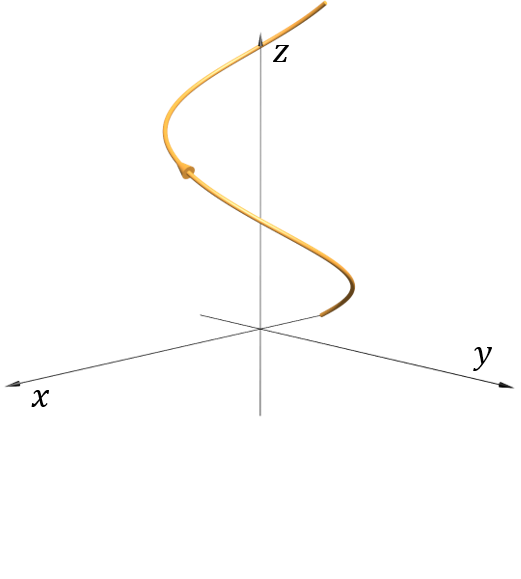}} \hspace{0.04in} \subfigure[]{ \label{004b}
\includegraphics[width=1.6in]{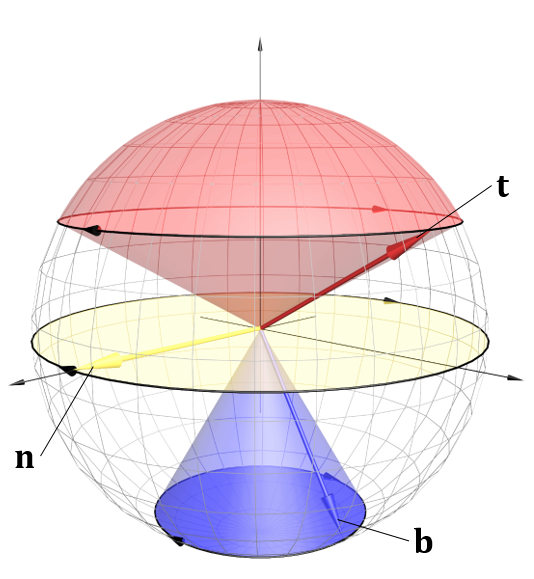}} \caption{(a) One period of an uniform helix with the pitch angle $\Theta$. (b) The rotations of the Frenet frame and the corresponding solid angles during the propagation along the helix.} \label{004}
\end{figure}

\subsection{Transverse-spin-dependent Hall effect}

Alternatively, the transverse-spin-dependent geometrical phase phenomena can be described by a more abstract and general formalism, that is the Berry connection within the Berry phase formalism. As in the situation of the optical Berry phase $\Phi_{\parallel}={\sigma_{\parallel}\int\tau ds}$, the Berry connection ${\bf A}_{\perp}$ and the corresponding Berry curvature ${\bf F}_{\perp}$ associated with the transverse spin are defined by \cite{Bliokh2008PRL}
\begin{eqnarray}\label{BerryFormalism}
  {\bf A}_{\perp}({\bf k})={\bf n}\cdot(\nabla_{\bf k}){\bf t},~~~~~~{\bf F}_{\perp}({\bf k})=\nabla_{\bf k}\times{\bf A}_{\perp}.
\end{eqnarray}
Importantly, these two geometrical quantities act as effective vector field and magnetic field in the ${\bf k}$-space, with $\sigma_{\perp}$ in Eq. (\ref{transversePhase}) playing the role of effective charge. By introducing the Berry phase formalism, we can calculate the geometrical phase Eq. (\ref{transversePhase}) through the Berry connection in the ${\bf k}$-space:
\begin{eqnarray}\label{transversePhaseA}
  \Phi_{\perp} &=& \sigma_{\perp}{\int_{C}{\bf A}_{\perp} \cdot d{\bf k}},
\end{eqnarray}
where $C$ is the contour of the evolution of ${\bf b}$ in the ${\bf k}$-space. In particular, for a closed contour the curvature-induced geometrical phase is determined by the solid angle enclosed by the contour.

Let us consider the propagation during one period of an uniform helix as a simple example (see Fig. \ref{004a}). The Berry connection and Berry curvature for such propagation can be easily obtained in spherical coordinates $(k, \theta, \phi)$:
\begin{eqnarray}\label{BerryC}
  {\bf A}_{\perp}({\bf k})=\frac{1}{k}(0,0,1),~~~~~~{\bf F}_{\perp}({\bf k})=\frac{1}{k^2}(\cot \Theta,0,0).
\end{eqnarray}
After the propagation, the transverse-spin-dependent geometrical phase can be calculated by Eqs. (\ref{transversePhase}) or (\ref{transversePhaseA}). Blow are the final result:
\begin{eqnarray}\label{GeometricalPhase0}
  \Phi_{\perp} &=&  \sigma_{\perp}2\pi \sin \Theta,
\end{eqnarray}
where $\Theta$ is the angle between the tangent vector and $z$ axis. Alternatively, one can determine this geometrical phase by calculating the solid angle subtended by the trajectory of ${\bf b}$ (see Fig. \ref{004b}),
\begin{eqnarray}\label{GeometricalPhase1}
  \Phi_{\perp}^B &=& -\sigma_{\perp}\Omega_{\perp}(C)= -\sigma_{\perp}\int_{0}^{2\pi}\int_{\pi/2+\Theta}^{\pi}\sin \theta d\theta d\phi \nonumber\\
   &=&\sigma_{\perp}2\pi(\sin\Theta- 1),
\end{eqnarray}
where $\Omega_{\perp}(C)$ is the solid angle enclosed by the contour $C$. Equation. (\ref{GeometricalPhase1}) coincides with Eq. (\ref{GeometricalPhase0}) up to $-\sigma_{\perp}2\pi$ difference caused by the rotation of the $\phi$-coordinates \cite{Bliokh2009JOA, Bliokh2015Natpho}. Passingly, the intriguing equivalence relation between the integration $\int\kappa ds$ and the solid angle $\Omega_{\perp}(C)$ can be alternatively demonstrated by referring to the argument provided in Ref. \cite{Haldane:86} by F. D. M. Haldane.

As we know, the geometrical phase and the spin Hall effect are two reciprocal phenomena \cite{Bliokh2009JOA}. For example, the spin Hall effect of light, which is associated with the optical Berry phase $\Phi_{\parallel}={\sigma_{\parallel}\int\tau ds}$, is described by \cite{Bliokh2009JOA}
\begin{eqnarray}\label{SHE0}
  \delta\dot{\bf r}_{\parallel} &=& -\sigma_{\parallel}k_0^{-1}\dot{\bf k}
\times{\bf F}_{\parallel} =-\sigma_{\parallel}k^{-1}\kappa{\bf b},
\end{eqnarray}
where ${\bf F}_{\parallel}$ is the effective magnetic field associated with the longitudinal spin of light. Thus, one can readily obtain a similar expression for the spin Hall effect associated with the transverse-spin-dependent geometrical phase (\ref{transversePhase}) by replacing the longitudinal quantities with transverse ones in the above equation. The result is
\begin{eqnarray}\label{SHE1}
  \delta\dot{\bf r}_{\perp} &=& -\sigma_{\perp}k_0^{-1}\dot{\bf k}
\times{\bf F}_{\perp} =\sigma_{\perp}k^{-1}\tau{\bf b}.
\end{eqnarray}
This quantity represents the transverse deflection of a light beam propagation along a curved path due to the presence of the transverse spin of light. This transverse-spin-dependent Hall effect is of the same magnitude comparing with the longitudinal one, and thus can be observed through optical experiments. In fact, the usual and transverse-spin-dependent geometrical phase phenomena correspond to different adiabatic approximations. Mathematically, the difference relates to what spin-1 matrix in Eq.~(\ref{tangentEq}) to be diagonalized, and the terminology ``adiabatic'' indicates rejecting the off-diagonal terms in Eq.~(\ref{tangentEq}) after the diagonalization \cite{BLIOKH200513}. Physically, the difference and ``adiabatic'' relate to the adiabatic transports of the circular polarizations or transverse circular polarizations. As we show in Sec.~\ref{GaugeStructure}, in the proper adiabatic approximation, the evolution of light can be adequately described by an U(1) gauge field. As the result, Eqs.~(\ref{SHE0}) and (\ref{SHE1}) are valid in different adiabatic approximations in terms of the Berry phase formalism. In conclusion, when transverse circular polarizations adiabatically transport along a curved path with nonvanishing torsion, a transverse-spin-dependent transverse deflection of the light beam occurs. This is the transverse-spin-dependent Hall effect of light.


\section{Gauge structure}\label{GaugeStructure}
Within the Berry phase formalism, the usual and transverse-spin-dependent geometrical phase phenomena are associated with different three-dimensional U(1) gauge fields: $ {\bf A}_{\parallel}({\bf k})={\bf b}\cdot(\nabla_{\bf k}){\bf n}$ and $ {\bf A}_{\perp}({\bf k})={\bf n}\cdot(\nabla_{\bf k}){\bf t}$ (see Ref.~\cite{Bliokh2009JOA} and Eq.~(\ref{BerryFormalism})). These two different U(1) gauge fields in the momentum space correspond to the torsion and curvature of the curve, respectively. As shown in Eq.~(\ref{tangentEq}), they are united in one wave equation.

Here, we discuss the effective gauge terms in Eq.~(\ref{tangentEq}) from the point of view of the gauge theory, and show that they can be described with respect to one group: the SO(3) group. To start with, we replace the Frenet frame $\{{\bf t, n, b}\}$ with an arbitrary triad $\{{\bf q}_i(s)\}$. The angular velocity matrices that describe the rotations of the triad along the curve are defined as follows
\begin{eqnarray}
\Omega_{ij}&=&{\bf q}_i\cdot\partial_s{\bf q}_j,
\end{eqnarray}
where $\{{\bf q}_i(s)\}$ are the three bases of the triad. Consequently, the angular velocity vector is defined in terms of these matrices as
\begin{eqnarray}\label{AngularVelocity}
  \Omega_i &=& -\frac{1}{2}\epsilon_{ijk}\Omega_{jk},
\end{eqnarray}
where $\epsilon_{ijk}$ is the three-dimensional Levi-Civita symbol. As an example, the angular velocity vector of the Frenet frame is
\begin{eqnarray}\label{AngularVelocityInFrenentFrame}
  {\bf \Omega}_{F} &=& \tau {\bf t} + \kappa {\bf b}.
\end{eqnarray}
Consider an arbitrary local rotation for the triad $\{{\bf q}_i\}$, which is represented by the local rotation angle ${\bm \theta}(s)$. The local rotation is given by
\begin{eqnarray}\label{SO3Transformation}
  {\bf q}_i &\longrightarrow& (e^{i\hat{\bf S}\cdot{\bm \theta}(s)})_{ij}{\bf q}_j.
\end{eqnarray}
As the result, the angular velocity matrix transforms as an SO(3) gauge connection
\begin{eqnarray}\label{TransformationOfAngularVelocity}
  \Omega_{ij} &\longrightarrow& \big[ (e^{i\hat{\bf S}\cdot{\bm \theta}})_{ik}{\bf q}_k\big]\partial_s\big[ (e^{i\hat{\bf S}\cdot{\bm \theta}})_{jl}{\bf q}_l\big] \nonumber\\
  &&=  (e^{i\hat{\bf S}\cdot{\bm \theta}})_{ik}\Omega_{kl} (e^{i\hat{\bf S}\cdot{\bm \theta}})_{jl}+(e^{i\hat{\bf S}\cdot{\bm \theta}})_{ik}\partial_s (e^{i\hat{\bf S}\cdot{\bm \theta}})_{jk}. \nonumber\\
\end{eqnarray}
Note that the torsion and curvature are geometrical quantities which can be directly measured, the term, $(e^{i\hat{\bf S}\cdot{\bm \theta}})_{ik}\partial_s (e^{i\hat{\bf S}\cdot{\bm \theta}})_{jk}$, which arises in the transformation between the Frenet frame and an arbitrary triad can be recognized as the ``pure gauge term''. The ``pure gauge term'' means that there is no observable effect associated with it.

Under the SO(3) rotaion (\ref{SO3Transformation}), the transformation for the electric field components corresponding to the triad is
\begin{eqnarray}\label{TransformationOfElectricField}
E_i
\longrightarrow  (e^{i{\bf \hat{S}}\cdot{\bm \theta}(s)})_{ij}E_j.
\end{eqnarray}
The transformations (\ref{TransformationOfAngularVelocity}) and (\ref{TransformationOfElectricField}) suggest the minimal coupling between the electric field and the SO(3) gauge connection
\begin{eqnarray}\label{Substituting}
  \partial_s {\bf E} &\longrightarrow& (\partial_s-i{\bf \Omega}\cdot\hat{\bf S} ){\bf E}.
\end{eqnarray}
In fact, note that $(\hat{S}_i)_{jk}=-i\epsilon_{ijk}$, it can be easily verified that
\begin{eqnarray}
&&(\partial_s-i{\bf \Omega}\cdot\hat{\bf S} ){\bf E}\longrightarrow (e^{i{\bf \hat{S}}\cdot{\bm \theta}(s)})[(\partial_s-i{\bf \Omega}\cdot\hat{\bf S} ){\bf E}].
\end{eqnarray}
Actually, in virtue of Eqs.~(\ref{AngularVelocityInFrenentFrame}), one can readily find that Eq.~(\ref{Substituting}) coincides with $(\partial_s-i\tau\hat{S}_1-i\kappa\hat{S}_3)$ in Eq.~(\ref{tangentEq}). The above argument demonstrate the emergence of a geometry-induced SO(3) gauge field during the curved propagation of light with nonvanishing longitudinal component.

In the spirit of the spin gauge field theory \cite{BLIOKH200513}, an U(1) gauge field arises in the adiabatic approximation \cite{Lai2018PRA}. The adiabatic approximation is associated with the diagonalizations of the three spin-1 matrices. Here, we take the diagonalization of $\hat{S}_3$ as an example (the case of $\hat{S}_1$ is presented in Ref.~\cite{Lai2018PRA}). The diagonalization can be achieved by the transformation (\ref{Diagonalization}). After the diagonalization, neglecting the off-diagonal terms leads to the reduction of Eq.~(\ref{tangentEq}) to Eq.~(\ref{tangentEqT=0,1}). Apparently, neglecting the off-diagonal terms corresponds to the adiabatic transports of the transverse circular polarizations. In this adiabatic approximation, the local rotation (\ref{SO3Transformation}) reduces to an SO(2) rotation
\begin{eqnarray}
  \left(
    \begin{array}{c}
      {\bf q}_1 \\
      {\bf q}_2 \\
    \end{array}
  \right)\rightarrow  e^{i\hat{\sigma}_s\theta(s)}\left(
    \begin{array}{c}
      {\bf q}_1 \\
      {\bf q}_2 \\
    \end{array}
  \right),
\end{eqnarray}
where $\hat{\sigma}_s=\left(
    \begin{array}{cc}
      0 & -i \\
      i & 0 \\
    \end{array}
  \right)$.
For simplicity, here we have neglected the trivial dimension ${\bf q}_3$. The transformations for the corresponding SO(2) gauge connection and the electric field becomes
\begin{eqnarray}
  \Omega_3 &\longrightarrow&  \Omega_3+\partial_s\theta,
\end{eqnarray}
and
\begin{eqnarray}
  E^{\sigma_{\perp}} &\longrightarrow& e^{i\sigma_{\perp}\theta(s)} E^{\sigma_{\perp}},
\end{eqnarray}
respectively, where $E^{\sigma_{\perp}}$ denote the two opposite transverse circular polarizations (see Eq.~(\ref{Diagonalization})). Consequently, an effective one-dimensional U(1) gauge field, note that SO(2) group is isomorphic to U(1) group, arises in the coordinate space. Within the Berry phase formalism, it corresponds to the three-dimensional U(1) gauge field, ${\bf A}_{\perp}({\bf k})$, in the momentum space.

\section{Conclusions}\label{conclusions}
In the present paper, we investigate the effective dynamics of light with nonvanishing longitudinal component by using the thin-layer approach. As the result, we obtain an effective equation (\ref{tangentEq}) which describes the curved propagation of light exhibiting longitudinal component. In addition to the torsion-induced term \cite{Lai2018PRA}, Eq.~(\ref{tangentEq}) contains a curvature-induced effective gauge term. This curvature-induced effective gauge term brings about a novel geometrical phase which is associated with the transverse spin along the binormal direction. Furthermore, we show that the curvature-induced effective gauge term is responsible for the nonadiabatic polarization changes of light \cite{berry1987interpreting,Lai2018PRA}. Imitating the reciprocal relation between the optical Berry phase and the spin Hall effect of light, we calculate the transverse-spin-dependent Hall effect of light. In contrast to the spin Hall effect of light \cite{Bliokh2008}, the novel Hall effect induced by the torsion is associated with the transverse spin along the binormal direction of the curve. Interestingly, we find that the usual (longitudinal-spin-dependent) and the novel (transverse-spin-dependent) geometrical phase phenomena are associated with different geometry-induced U(1) gauge fields in different adiabatic approximations. These gauge fields are united in Eq.~(\ref{tangentEq}) by SO(3) group in the nonadiabatic case. Therefore, equation (\ref{tangentEq}) provides a more adequate description of the nonadiabatic evolution of light.

As discussed, the transverse-spin-dependent geometrical phase effects have purely geometrical origin like the longitudinal-spin-dependent ones. It therefore provides new possibilities to control the spin and orbital degrees of freedom of light via the geometries. For example, recently the transverse-spin-dependent geometrical phase has been employed to generate the SOI between the intrinsic orbit AM and the transverse spin of light \cite{Shao2018}. We believe that the present research could benefit the rapidly growing research of the transverse spin of light.

\acknowledgments
We thank Prof. Fan Wang for very helpful discussions about the gauge theory. This work was supported in part by the National Natural Science Foundation of China (under Grants No. 11690030, No. 11535005, No. 11625418, and No. 51721001), the Fundamental Research Funds for the Central Universities (under Grant No. 020414380074). Y.-L.W. was funded by the Natural Science Foundation of Shandong Province of China (Grant No. ZR2017MA010), and Linyi University (Grant No. LYDX2016BS135).

\appendix*
\section{From the point of view of the Coriolis effect}
As discussed above, the geometrical phases of light is straightforward consequences of the Maxwell's equations, and is associated with the parallel transport of the electric field vector. Alternatively, they can be derived through a different way. Note that an arbitrary rotation with respect to a laboratory coordinate frame is described by a precession of the triad attached to the curve with the angular velocity vector (\ref{AngularVelocity}) \cite{Bliokh2009JOA}.
Analogous to the Coriolis effect in classical mechanics \cite{Goldstein2014}, some inertia terms arise in the Helmholtz equation via the following substitution \cite{Lipson:90,Bliokh2009JOA}:
\begin{eqnarray}\label{Coriolis}
  \partial_t {\bf E} \longrightarrow \partial_t {\bf E} + \frac{c}{n}{\bf \Omega}\times{\bf E},
\end{eqnarray}
where the angular-velocity-induced term is the Coriolis term in this situation. Since the time $t$ is associated with the arc-length $s$ by the wave velocity $c/n$, the Coriolis term can be alternatively introduced through the following way:
\begin{eqnarray}\label{Coriolis1}
  \partial_s {\bf E} &=&  \partial_s (E_t{\bf t}+ E_n{\bf n}+ E_b{\bf b}) \nonumber\\
  &=&  (\partial_s E_t){\bf t}+ (\partial_s E_n){\bf n}+ (\partial_s E_b){\bf b}+{\bf \Omega}\times{\bf E}.
\end{eqnarray}
Importantly, the last row in the above equation can be rewritten in the following form:
\begin{eqnarray}
  (\partial_s-i{{\bf\Omega}\cdot\hat{\bf S}})\left(
                                               \begin{array}{c}
                                                E_t \\
                                               E_n \\
                                               E_b \\
                                               \end{array}
                                             \right),
\end{eqnarray}
which coincides with Eq.~(\ref{tangentEq}). Note that we have used a special triad -- the Frenet frame -- rather than an arbitrary triad in the above derivations. In the Frenet frame, the rotation of ${\bf b}-{\bf t}$ plane vanishes, and thus the second spin-1 matrix $\hat{S}_2$ does not appear in Eq.~(\ref{tangentEq}). In conclusion, we have demonstrated the equivalence between the effective gauge terms in Eq.~(\ref{tangentEq}) and the Coriolis terms here. It is obvious that the derivation of the Coriolis terms is much simpler. Nevertheless, like the Coriolis effect in classical mechanics, the introducing of the Coriolis terms here is somewhat artificial. Most importantly, the Coriolis terms are irrelevant to the spatial distribution of the electric field, which means that they can not describe the geometrical phase phenomena associated with the intrinsic orbital AM. On the other hand, as we have shown in Ref.~\cite{Lai2018PRA}, the orbital geometrical phase can also be directly derived from the Maxwell's equations by using the same approach presented in Sec.~\ref{effectiveEq}. Physically speaking, once we replace the ordinary differential with the covariant differential, the parallel transport of the electric field are included in the Maxwell's equations. In particular, the parallel transport of the vortex is also included as the polarization.

\normalem
\bibliographystyle{apsrev4-1}

\end{document}